\documentclass{article}
\usepackage[x11names]{xcolor}
\usepackage{graphicx}
\usepackage{dcolumn}
\usepackage{bm}
\usepackage{upgreek}
\usepackage{adjustbox}
\usepackage[utf8]{inputenc}
\usepackage[T1]{fontenc}
\usepackage[colorinlistoftodos]{todonotes}
\usepackage{mathptmx}
\usepackage{etoolbox}
\def\Mu{\upmu}
\usepackage{geometry}
\begin{document}

\title{Spin-wave spectra in antidot lattice with inhomogeneous perpendicular magnetocrystalline anisotropy}

\author{M. Moalic, M.Krawczyk and M. Zelent}
\date{Institute of Spintronics and Quantum Information, Faculty of Physics, Adam Mickiewicz University, Poznan, Poland}

\maketitle

\begin{abstract}
Magnonic crystals are structures with periodically varied magnetic properties that are used to control collective spin-wave excitations. With micromagnetic simulations, we study spin-wave spectra in a 2D antidot lattice based on a multilayered thin film with perpendicular magnetic anisotropy (PMA). We show that the modification of the PMA near the antidot edges introduces interesting modifications to the spin-wave spectra, even in a fully saturated state. In particular, the spectra split in two types of excitations, bulk modes with amplitude concentrated in a homogeneous part of antidot lattice, and edge modes with an amplitude localized in the rims of reduced PMA at the antidot edges.  Their dependence on the geometrical or material parameters is distinct but at resonance conditions fulfilled, we found strong hybridization between bulk and radial edge modes. Interestingly, the hybridization between the fundamental modes in bulk and rim is of magnetostatic origin but the exchange interactions determine the coupling between higher-order radial rim modes and the fundamental bulk mode of the antidot lattice.  
\end{abstract}

\section{Introduction}
Magnonic crystals (MCs), analogous to photonic crystals that have a periodic modulation of the refractive index to control the propagation of electromagnetic waves, are magnetic structures with a periodicity of material properties relevant to spin-wave (SW) propagation.\cite{Gulyaev2001,Puszkarski03} Their main feature is the formation of bands of collective SW excitations and the separating them magnonic bandgaps, which is useful for applications to guide and control spin waves.\cite{Lenk2011,Krawczyk14,Nikitov2015,Rychly2015,Chumak2017} The basic types of MCs are artificial crystals formed by periodic arrangement of two different materials.\cite{Krawczyk14} Those can be two ferromagnetic materials, like in bicomponent MCs,\cite{Tacchi12} ferromagnetic and non-magnetic materials like in an array of ferromagnetic dots in nonmagnetic matrix\cite{Kruglyak2010,Tacchi2011,Saha2013}, or the inverse structures,\cite{Kostylev2008,Mandal2013} i.e., an array of holes in a ferromagnetic matrix, antidot lattices (ADLs).\footnote{Recently, the term of MC has been extended to the lattice of atomic spins, which form the spin-wave band structure at large, up to THz, frequency range.\cite{Zakeri2020}} 

Periodic modulation of SW-relevant parameters can be introduced not only in the patterning process\cite{Martin2003,Lau2011} but also at a later stage. For instance, by ion irradiation of the ferromagnetic film or multilayers, to modify the magnetic properties on the exposed areas.\cite{Carter1982,Urbanek2018,Fassbender2009,Obry2013,Wawro2018} Recently, another promising idea has been explored, that is, the formation of periodicity by regular change of the magnetization orientation in a homogeneous thin film.\cite{Yu2021} The basic examples are the periodic stripe domains, which can be considered as 1D MCs\cite{Banerjee2017,Szulc2022} or skyrmion lattices, which also form a kind of 2D MC.\cite{Diaz2020,Chen2021,Takagi2021,Bassotti2022} These structures have an important property; their magnetization structure is reconfigurable and sensitive to the external bias magnetic field; in this way, the magnonic band structure can be programmed after the manufacturing of the device, depending on the actual needs.\cite{Yu2021} Nevertheless, the structures combining both patterning and magnetization texture are not yet well explored, in particular in the context of MCs and magnonics. 

In recent studies, ADL based on multilayers with perpendicular magnetic anisotropy (PMA) has been indicated as a patterned system hosting periodic magnetization texture with interesting properties and potential applications.\cite{Pal2014,Pan2020c,Mantion2022} The results of time-resolved magneto-optical Kerr effect (TR-MOKE) microscopy measurements of SW excitations in ADL based on [Co/Pd]$_8$ multilayers were interpreted with micromagnetic simulations.\cite{Pal2014} To explain low frequency mode in the spectra, the authors assumed that around the holes there is a rim with modified magnetic properties formed during the patterning process, i.e., Ga ions used in focused ion beam (FIB) in the patterning process penetrate larger area than the antidots. In the rims around the antidots, the dose is sufficient to disturb the interfaces between Co and Pd, and so, the PMA, which originates from the interfaces in the multilayer. The absence or reduced PMA in the rims results in the in-plane alignment of the magnetization in the rims at remanence.\cite{Pan2020c} The MC composed of two areas of different magnetization orientations, out-of-plane in the bulk and in-plane in the rims, makes the SW spectra complex, with bulk modes localized in the ferromagnetic matrix, as in standard ADLs, and the additional modes localized in the rims, with possible hybrid excitations. Such structures open the prospects for magnonic band structure with sub-bands sensitive to external stimuli in different ways, e.g., different dependences on a magnetic field, the property, which can be exploited for developing new phenomena and applications. 

In this paper, we study the SW spectra in the ADL lattice based on multilayer thin film with PMA with the rims around the antidots, where the PMA is reduced. In particular, we study the dependence of the spectra on the rim anisotropy value, rim width and antidot diameter under a perpendicular magnetic field, fully saturating the sample. We show that the variation of these properties affects mainly the edge-localized modes, but additionally may introduce or control a hybridization between the radial modes of the rim and the bulk excitations of the ADL. This offers a possibility to control collective SWs by tuning magnetic properties only in selected areas. 

The paper is organized as follows. In the first section, we introduce the structure and the micromagnetic model used in the simulations. In Sec.~\ref{Sec:Results} we present the simulation results of SW spectra in dependence on the anisotropy strength in the rim, antidot diameter and finally rim width, and interpret the obtained results. In the last Sec. IV we summarise the results. 

\section{Structure and methodology}
\label{Sec:Methods}

\begin{figure}[htbp]
\includegraphics[width=1.0\linewidth]{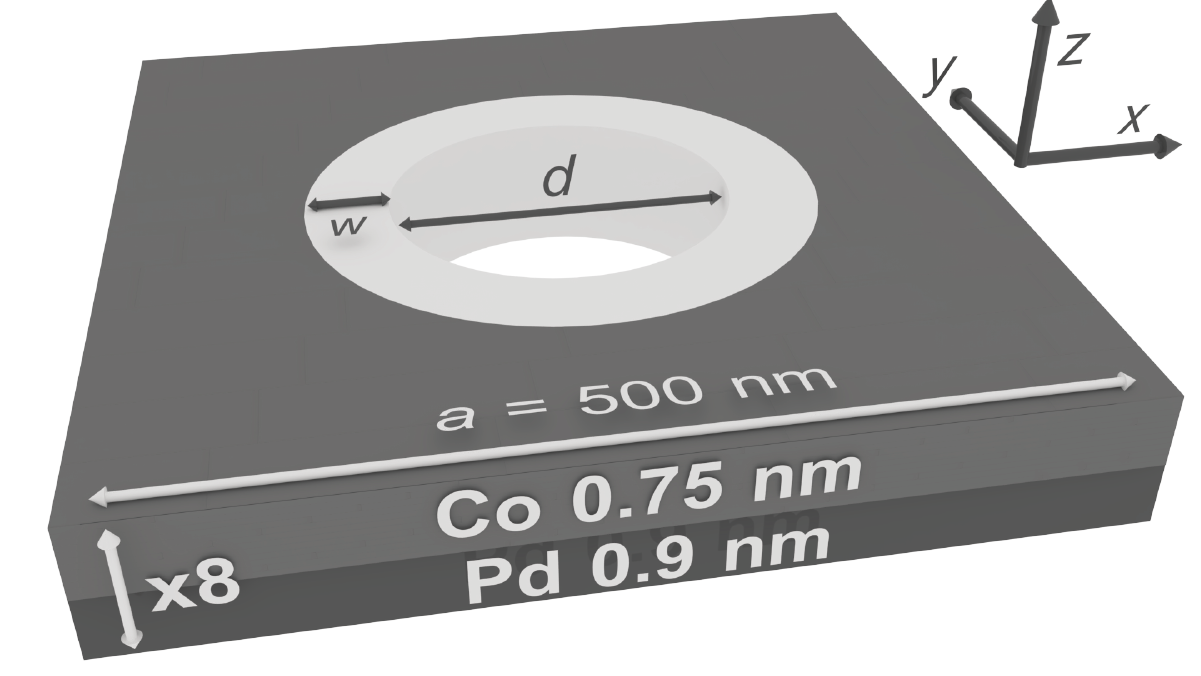}
\caption{
Schematic illustration of the investigated structure showing the Co/Pd supercell of ADL. Note that the figure is not to scale. The white color around the hole represents the rim with the reduced PMA. 
}
\label{geom}
\end{figure}

For our numerical investigations, we selected a structuralized Co/Pd multilayer with high PMA (with the anisotropy constant $K_{\mathrm{u,bulk}}$) and reduced anisotropy around the antidots, already investigated in Ref.~\cite{Pan2020c}. The schematic of the unit cell of the studied structure is shown in Fig.~\ref{geom}.
The structure is made up of 8 repetitions of Co (0.75 nm thickness) and Pd (0.9 nm) bilayers with a total thickness of 13.2 nm. Around the antidot of diameter $d$ is a rim of width $w$ with reduced PMA, $K_{\mathrm{u,rim}}$. In our study, we will change these three parameters. The antidots form a square lattice on the $(x,y)$ plane with a lattice constant $a = 500$ nm. To maintain the same static magnetization configuration of the system, i.e. to maintain a full saturation of magnetization in the out-of-plane direction, we perform numerical simulations using a sufficiently strong external magnetic field $\mu_0 H_{\textrm{ext}}=1$ T directed along the $z$ axis. 

We use micromagnetic simulations with Mumax3 software,~\cite{mumax_2014,Leliaert2018} which solves the Landau-Lifshitz-Gilbert equation for magnetization vector $\textbf{M}$:
\begin{equation}
 \frac{\textrm{d}\mathbf{m}}{\mathrm{d}t}=\gamma \Mu_0 
 \frac{1}{1+\alpha^{2}} (\mathbf{m} \times \mathbf{H}_{\mathrm{eff}}) + 
 \alpha\Mu_0 \left( \mathbf{m} \times 
 (\mathbf{m} \times \mathbf{H}_{\mathrm{eff}}) \right),
\end{equation}
where $\textbf{m} = \textbf{M} / M_{\mathrm{S}}$ is the normalized magnetization, $M_{\mathrm{S}}$ is the magnetization saturation, $\textbf{\textrm{H}}_{\mathrm{eff}}$ is the effective magnetic field acting on the magnetization, $\gamma=-187$ rad GHz/T is the gyromagnetic ratio, $\alpha$ is the damping constant.
The following components were considered in the effective magnetic field $\textbf{H}_{\mathrm{eff}}$: demagnetizing field $\textbf{\textrm{H}}_{\mathrm{d}}$, exchange field $\textbf{\textrm{H}}_{\mathrm{exch}}$, uniaxial magnetocrystalline anisotropy field $\textbf{\textrm{H}}_{\mathrm{Ku}}$, and external magnetic field, thermal effects were neglected. Thus, the effective field  is expressed as:
\begin{equation}
  \textbf{H}_{\mathrm{eff}} =
  \textbf{H}_{\mathrm{d}} + \textbf{H}_{\mathrm{exch}} + \textbf{H}_{\mathrm{ext}} + \textbf{H}_{\mathrm{Ku}} +\textbf{h}_{\mathrm{mf}},
\end{equation}
where the last term, $\textbf{h}_{\mathrm{mf}}$ is a microwave magnetic field used for SW excitation. The exchange and anisotropy fields are defined as:
\begin{equation}
  \textbf{H}_{\mathrm{exch}} = \frac{2A_{\mathrm{ex}}}{\mu_0 M_{\mathrm{S}}} \Delta \textbf{m},\;
  \textbf{H}_{\mathrm{Ku}} =
\frac{2K_{\mathrm{u,bulk}}}{\mu_0 M_{\mathrm{S}}} 
\left( \textbf{u} \cdot \textbf{m} \right) \textbf{u},
\label{Eq:Fields}
\end{equation}
where $\textbf{u} $ is the unit vector that indicates the direction of anisotropy, $A_{\textrm{ex}}$ is the exchange constant.

To simplify the micromagnetic simulations of the multilayered ADL system, we used the effective thickness approach, which involves the effective, experimentally measured, values of the magnetic parameters for a single discretization through the thickness of the magnetic multilayer.~\cite{Pal2011d} 
In the simulations, we used a cell size of $500/512 \approx 0.97$ nm for the $x$ and $y$ axis, and 13.2 nm for the $z$ axis. To mimic a periodic structure, we used 32 repetitions of Mumax3 periodic boundary conditions, along the $x$ and $y$ axes.  
The effective magnetic parameters for the Co/Pd multilayer are taken from Ref.~\cite{Pan2020c}, these are $K_{\mathrm{u,bulk}} = 4.5\times 10^5$ J/m$^3$, $M_{S} = 0.81 \times 10^6$ A/m, $A_{\textrm{ex}} = 1.3 \times 10^{-11}$ J/m. We used a low damping constant $\alpha = 1 \times 10^{-7}$ to simulate sharp SW spectra. 

\begin{figure*}
\includegraphics[width=1.0\textwidth]{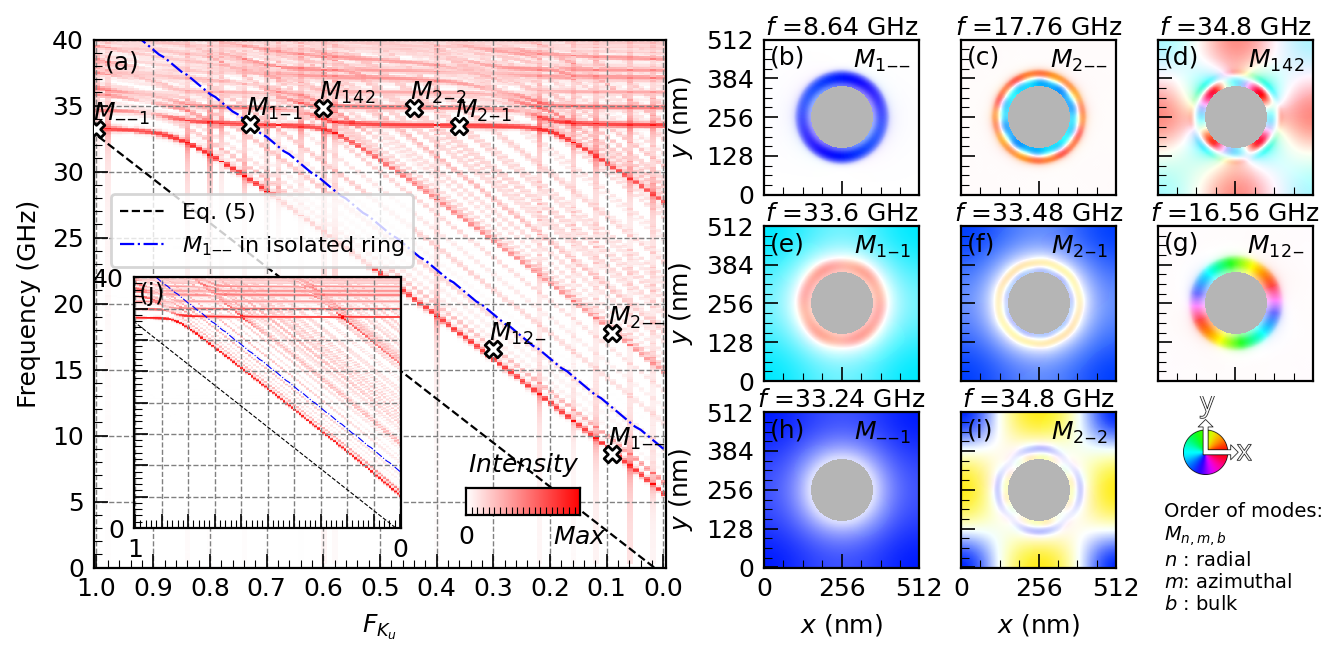}
\caption{(a) SW spectrum dependent on the reduced anisotropy strength $F_{K_\textrm{u}}$. The intensity of the SW spectra has been presented in a logarithmic scale colormap. 
The black-dashed line represents the analytically calculated ferromagnetic resonance frequency of the uniform thin film, Eq.~(5).
The blue-dashed-dotted line represents the resonance frequency of the fundamental mode in the isolated ring. The markers indicate the SW modes illustrated in the right panels (b-i). Here, the color saturation indicates the mode intensity and the hue indicates the phase of the mode, according to the diagram in the right bottom corner.  Inset (j) presents analogous dependence to panel (a), but with rim-matrix exchange interactions disabled.  
\label{fig:rim_ku}}
\end{figure*}

To obtain full SW spectra first, we relax the system, then we implement a special distribution of the external microwave magnetic field for SW excitation, $\textbf{h}_{\mathrm{mf}}$. Using the Voronoi tessellation algorithm, we divided the ADL unitcell area into small random sub-areas, about 30 nm wide each, effectively acting as SW point sources (nanoantennas). Each of these nanoantennas was assigned the same amplitude of a microwave magnetic field, but a random phase offset was used in the system's excitation by an in-plane magnetic field, along the diagonal of the ADL. To excite SWs in a wide frequency range we used a $sinc$ function to apply a time modulation of the microwave field, which has a maximum amplitude of 0.5 mT, a cutoff frequency of 40 GHz, and a sampling rate of 8.3 ps. The magnetic field pumps the system for 1 ns and stops before any data are recorded. The time-resolved, space-dependent magnetization was then saved over 8.3 ns and processed by the Fast Fourier transform (FFT) to get 2D visualization of all SW modes. To obtain the SW spectra, we applied a Hanning window to both the space and time dimensions of the data before using the FFT. We then found for each frequency, the cell in the system that had the highest amplitude and used it as the FFT amplitude in the spectra. This approach to calculate the SW spectra has the benefit of not favoring the extended SW modes, as is in the case of homogeneous microwave field.  

To refer to a particular SW mode, we use the $M_{mnb}$ notation, where we identify the radial $n$, azimuthal $m$, and bulk $b$ numbers, which represent the order of the modes in the rim along the radial and azimuthal directions, and in the bulk, respectively. 
For bulk, by the mode order, we took the number of nodal points along the line connecting neighboring antidots minus one. For instance, a homogeneous in space mode, a fundamental mode, which has a pinned magnetization at the edges of the antidots, we will have the order number $b=1$. A bulk mode with one more nodal point between antidots, i.e., in total three nodes, is a second-order bulk mode with $b=2$. When a number is not applicable, we replace it with the '-' symbol. The radial mode order is defined as a number of amplitude quantizations in the rim along the radial direction, where $n=1$ means a fundamental mode with only partial pinning at the rim edges and no nodes between, then $n=2$ represents a mode with 1 node between the rim edges. For azimuthal modes, the value of the number $m$ defines the number of full phase rotations along the circumference, where $m=1$ means a  $2 \pi$ phase change along the circumference of the rim. 

\section{Results}
\label{Sec:Results}
Initially, we simulate the ADL with an antidot diameter of $d=200$ nm and a rim width of $w=50$ nm depending on the anisotropy value in the rim.
As the $\textbf{\textrm{H}}_{\mathrm{ani}}$ is parallel to $\textbf{\textrm{H}}_{\mathrm{ext}}$, a local reduction of the anisotropy field should result in a decrease of the resonant frequencies of SWs confined in these areas. Indeed, Fig.~\ref{fig:rim_ku}  (a) shows the dependence of the SWe spectrum on the reduced magnetocrystalline anisotropy factor $F_{K_\textrm{u}}$: 
\begin{equation}
 F_{K_\textrm{u}} = \frac{K_{\mathrm{u,rim}}}{K_{\mathrm{u,bulk}}},
\end{equation}
which reveals two distinct groups of modes. One group with frequencies linearly decreasing as $F_{K_\textrm{u}}$ does, and the second one, only weakly dependent on the anisotropy in the rim, which corresponds to the ADL modes in the rim and bulk, respectively. The representative eight modes are marked in Fig.~\ref{fig:rim_ku} (a) and their in-plane spatial profiles are shown in Fig.~\ref{fig:rim_ku} (b)-(i). 

\begin{figure*}[htbp]
\includegraphics[width=1.0\textwidth]{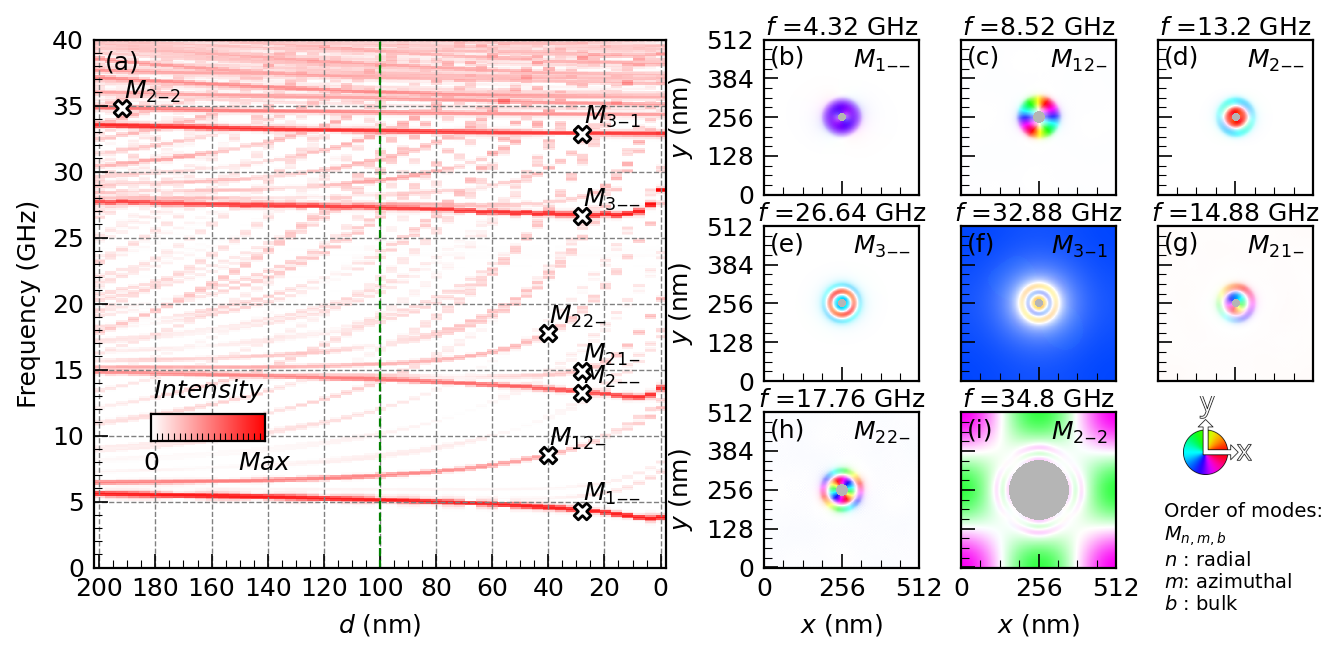}
\caption{ 
(a) The spin-wave spectrum in dependence on the antidot diameter $d$. The green-dashed line represents the antidot diameter $d$=100 nm used in Fig.~2 and Fig.~4.
(b-i) Visualization of spin-wave modes for selected frequencies and antidot diameters marked in (a). The saturation indicates the intensity and the hue indicates the phase of the mode, according to the diagram in the right bottom corner. 
}
\label{fig:antidot_diameter}
\end{figure*} 

For an uniform distribution of anisotropy, $F_{K_\textrm{u}}=1$, the lowest frequency mode is a fundamental bulk mode $M_{ \mathrm{-} \mathrm{-}1}$ at the frequency 33.24 GHz, illustrated in Fig.~\ref{fig:rim_ku} (h). The modes of higher frequencies are the higher-order bulk modes. The reduction of the anisotropy in the rims leads to the transition of a homogeneous ADL to a structuralized bicomponent system with different magnetic parameters, and to isolate edge-localized radial and azimuthal modes. Therefore, with decreasing $F_{K_\textrm{u}}$, the frequencies of these two types of modes decrease, and crossover the bulk modes.
At some $F_{K_\textrm{u}}$ we observe strong hybridization, in particular at crossings of the fundamental bulk mode $M_{ \mathrm{-} \mathrm{-}1}$ at $F_{K_\textrm{u}}=$ 0.86, 0.59, and 0.21. The visualizations of modes presented in Fig.~\ref{fig:rim_ku} (e) and (f) shows that these hybridizations are between the mode $M_{ \mathrm{-} \mathrm{-}1}$ and the radial modes of the rim: $M_{1 \mathrm{-} \mathrm{-}}$, $M_{2 \mathrm{-} \mathrm{-}}$, and $M_{3 \mathrm{-} \mathrm{-}}$ (profile not shown). The first two radial modes are shown in Fig.~\ref{fig:rim_ku} (b) and (c), and their hybridizations with fundamental bulk mode in (e) and (f), respectively. Also, higher order bulk modes hybridize with the rim modes, although with smaller strength. For example, the hybridization of the second-order radial mode $M_{2 \mathrm{-} \mathrm{-}}$ with the second-order bulk mode $M_{ \mathrm{-} \mathrm{-}2}$, is shown in Fig.~\ref{fig:rim_ku} (i) as $M_{2 \mathrm{-}2}$. The hybridization of the modes $M_{ \mathrm{-} \mathrm{-}2}$ and $M_{14\mathrm{-}}$ is shown in Fig.~\ref{fig:rim_ku} (d), which clearly demonstrates the hybridization of the bulk modes also with the azimuthal modes of the rim.

At frequencies smaller than 33.24 GHz we see the parallel lines of different intensities. The intensive lines have already been identified as a radial modes of the rim; small intensity lines are azimuthal modes, for example, at 16.56 GHz is the $M_{12\mathrm{-}}$ mode presented in Fig.~\ref{fig:rim_ku} (g). 
The first-order radial mode in the rim $M_{1\mathrm{-}\mathrm{-}}$ is analogous to the bulk fundamental mode but restricted in space to the rim. Thus, the $M_{1\mathrm{-}\mathrm{-}}$ and $M_{\mathrm{-}\mathrm{-}1}$ are two fundamental excitations of our ADL. Their mode frequencies are influenced by the anisotropy, exchange, and magnetostatic fields. To analyze the role of individual interactions on the coupling between the modes, we make a comparison of the full spectra Fig.~\ref{fig:rim_ku} (a) with the spectra of simplified structures. 

\begin{figure*}[htbp]
\includegraphics[width=1.0\textwidth]{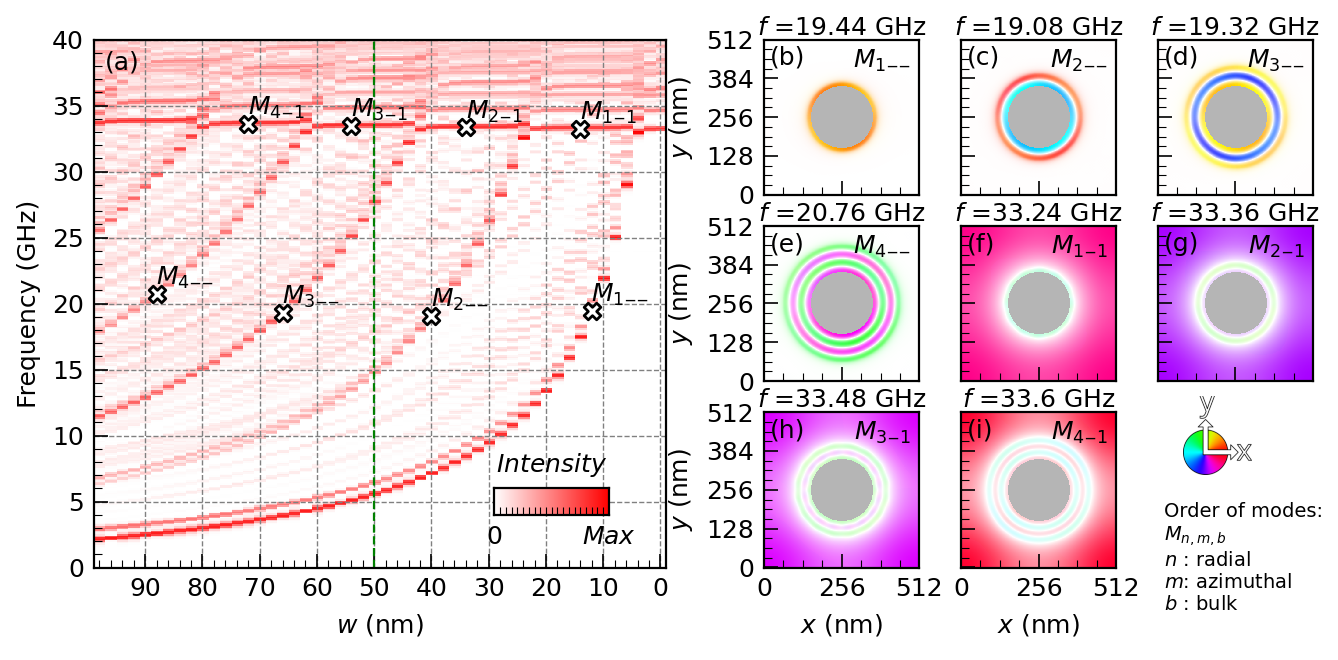}
\caption{ \label{fig:rim_width}
(a) Spin-wave spectrum in dependence on rim width, $w$. The green-dashed line represents the rim width used in previous investigations. (b-i) Visualization of spin-wave modes at selected frequencies and antidot diameters marked in (a). The color saturation indicates the mode intensity and the hue indicates the phase of the mode, according to the diagram in the right bottom corner. 
}
\end{figure*} 

To clarify the direct effect of the anisotropy field on the system, we calculate the ferromagnetic resonance frequency of a homogeneous film by varying the  anisotropy field. 
We use Kittel's formula:\cite{Kittel1948}
\begin{equation}
f_{0}=\frac{|\gamma|\mu_{0} (H_{\textrm{ani}} + H_{\mathrm{d}}+ {H}_{\mathrm{Ku}})}{2\pi},
\label{eq:kittel}
\end{equation}
with $f_{0}$ being the mode frequency, $H_{\mathrm{d}}=- M_{\mathrm{S}}$, and $H_{\mathrm{Ku}}$ as defined in Eq.~(\ref{Eq:Fields}). The function $f_0(F_{K_\textrm{u}})$ is shown with black-dashed line in Fig.~\ref{fig:rim_ku} (a).
When $F_{K_\textrm{u}}=1$, $f_0$ and the resonance frequency for the simulated ADL are only 400 MHz apart. This means that an ADL with an antidot diameter of $a=200$ nm at full saturation has a minor effect on the resonant frequency of the fundamental bulk mode. However, as $F_{K_\textrm{u}}$ is reduced, the fundamental mode smoothly transforms into a radial mode $M_{1\mathrm{-}\mathrm{-}}$ in the rim. At $F_{Ku}=0.84$ we observe a strong hybridization between both modes; see also the mode $M_{1\mathrm{-}1}$ in Fig.~\ref{fig:rim_ku} (e). With further reduction of anisotropy in the rim, the frequency difference increases, reaching a stable frequency separation of 6.20 GHz between $f_0$ and the $M_{1\mathrm{-}\mathrm{-}}$ mode. This difference between frequencies is mainly due to shape anisotropy of the rim, which rises the rim mode frequencies. 

If we compare the frequencies of the $M_{1\mathrm{-}\mathrm{-}}$ mode with those of the isolated ring  (obtained from independent micromagnetic simulations, keeping parameters of the rim), the green-dashed line in Fig.~\ref{fig:rim_ku} (a), we can see that the frequency of the fundamental mode of the rim is constantly lower by 3.59 GHz. This means that the magnetostatic field from the ADL bulk and a side contact of the rim with the ADL, effectively lowers the internal demagnetizing field in the rim, lowering its mode frequencies. 

To estimate the influence of the exchange interactions, in particular on the strength of the hybridizations and a frequency of the excitations, we performed additional micromagnetic simulations. Namely, we excluded the exchange interactions between the rim and bulk by introducing a 2 nm nonmagnetic spacer. The result is presented in Fig.~\ref{fig:rim_ku} (j). We notice that the frequency of the fundamental bulk mode $M_{\mathrm{-}\mathrm{-}1}$ at $F_{K_\textrm{u}}=0.85$ only slightly rises from 32.88 to 33.48 GHz. Thus, the hybridization strength has slightly decreased with the exchange decoupling of the rim from the bulk. For the hybridization of the $M_{\mathrm{-}\mathrm{-}1}$ mode with higher order radial modes, at $F_{K\textrm{u}}=0.59$, and 0.21, the frequency difference does not exceed 300 MHz in Fig.~\ref{fig:rim_ku} (j), indicating very weakly coupled modes, when the exchenge between rim and bulk is excluded.  This clearly indicates that these hybridizations are driven by the exchange interactions. Thus, we can conclude that the hybridization of fundamental bulk mode  with fundamental rim mode has a magnetostatic character, while the hybridizations with higher order radial modes have a strong exchange character.

For further analysis, we performed micromagnetic simulations of the ADL in dependence on the antidot diameter in the range from 200 to 0 nm (without antidot), with a 2 nm step; the other parameters are fixed, i.e., $F_{K\textrm{u}}=0$ and $w=50$ nm. The results are shown in Fig.~3 (a).
As expected, the diameter of the antidot only weakly affect the frequency of the fundamental bulk mode [see the mode $M_{3\mathrm{-}1}$ and Fig.~3 (f)] and the higher-order bulk modes [e.g., $M_{2\mathrm{-}2}$ and Fig.~3 (i)]. Unexpectedly, we do not observe   avoid-crossings as $d$ changes, although many crossings between different types of modes are visible. On the other hand, Fig.~3 (f) clearly indicates the hybrydisation occuring between the fundamental bulk mode and the radial rim modes. 

The frequency of the rim radial modes shows a slight decrease with decreasing $d$ [see, the modes $M_{1\mathrm{-}\mathrm{-}}$, $M_{2\mathrm{-}\mathrm{-}}$, and $M_{3\mathrm{-}\mathrm{-}}$ and their profiles in Fig. 3 (b), (d), and (e), respectively]. But at an antidot diameter of about 10 nm, the frequency of these modes starts to increase. It is caused by a change in the topology of the ADL geometry. The structure changes from an array of antidots to a full ferromagnetic layer. There are strong dipole interactions on opposite sides of the small antidot at this diameter range, which result in a change of the effective demagnetizing factors and an increase in mode frequency.\cite{Centala2019} 
The frequency of the second and higher-order azimuthal modes linearly increases at high rates, for example, the frequency of the second-order azimuthal mode $M_{12\mathrm{-}}$ and $M_{22\mathrm{-}}$ [Fig.~3 (c) and (h)] increases from 6.5 GHz at $d= 200$ nm up to 10 GHz at 20 nm antidots. This is due to  reduction in circumference of the antidot rim which contains the wavelength of the azimuthal mode to decrease, resulting in an increase in frequency. 

The other important parameter that affects the resonance spectrum of SWs is the width of the rim $w$. We performed micromagnetic simulations with a varied width of the reduced anisotropy region from 0 to 100 nm for a fixed antidot diameter ($d=200$ nm) (see, Fig.~4). The results show that the width of the rim significantly affects the frequency of radial and azimuthal modes [see, modes $n=1 \cdots 4$ and their profiles in Fig.~4 (b)-(e)], but having no significant effect on the frequency of the bulk modes (e.g., see the frequency of the fundamental bulk mode $M_{n\mathrm{-}1}$). As the rim area decreases, the frequency of azimuthal and radial modes in the rim increases, although at different rates. For example, the frequency of the mode $M_{1\mathrm{-}\mathrm{-}}$ increases from about 2.04 to 33.24 GHz, where the hybridization with the fundamental bulk mode $M_{\mathrm{-}\mathrm{-}1}$ occurs; see the hybridized mode $M_{1\mathrm{-}1}$ in Fig.~4 (f). Interestingly, there are hybridizations between the $M_{\mathrm{-}\mathrm{-}1}$ mode and the radial modes $M_{n\mathrm{-}\mathrm{-}}$ of the rim, but unexpectedly, the strength of these hybridizations increases with $n$.

\section{Conclusions}\label{Sec:Summary}
We have studied the SW spectrum in an ADL made of Pd/Co multilayers with PMA at full saturation induced by a PMA and out-of-plane oriented external magnetic field. A characteristic feature of the structure under investigation is a region around the antidots, a rim, in which the anisotropy was reduced. We have shown that in such a system there are extended bulk modes, and radial and azimuthal excitations. The last two are confined to the rim area and they are very sensitive to variation of the local rim parameters. In particular, the frequency of the azimuthal modes increases with decreasing antidot size, while the dependence of the radial modes is not monotonous in that case. Both types of SWs decrease the frequency with decreasing an anisotropy in the rim or with increasing a rim width. 

Interestingly, it is possible to achieve strong interaction between bulk modes of the ADL and the rim localized modes. Particularly strong hybridization has been found between the fundamental mode of ADL and radial excitation of the rim, reaching a value of 1.5 GHz. Importantly, such a coupling is realized by magnetostatic interaction between fundamental modes of the bulk and the rim, while mainly by the local exchange coupling between the rim area and the bulk of ADL for hybridizations with higher order radial modes. This opens the way for designing magnonic crystals with hybridized localized and extended modes, scaled to the deep nanoscale and properties controlled locally at the rim areas.
These interesting properties can be especially useful when designing multifunctional SW devices.

\section{Acknowledgments}
The research has received funding from the National Science Centre of Poland, Grant No.~UMO--2020/37/B/ST3/03936. The simulations were partially performed at the Poznan Supercomputing and Networking Center (Grant No.~398).

\bibliographystyle{abbrv}
\bibliography{sample}

\end{document}